\documentclass[aps,twocolumn,superscriptaddress,amsmath,amssymb,nofootinbib,nobibnotes,physrev,floatfix]{revtex4-2}
\pdfoutput=1 
\usepackage[utf8]{inputenc}
\usepackage[T1]{fontenc}
\usepackage[version=3]{mhchem}
\usepackage{graphicx}  
\usepackage{dcolumn}   
\usepackage{bm}        
\usepackage{units}
\usepackage[english]{babel}
\usepackage[dvipsnames]{xcolor}
\usepackage[
colorlinks=true,
urlcolor=RoyalBlue,
linkcolor=RoyalBlue,
citecolor=RoyalBlue,
]{hyperref}
\usepackage{vmargin}
\setpapersize{A4} \oddsidemargin20mm \textwidth170mm \topmargin10mm
\usepackage[type1]{libertine}
\usepackage{textcomp}
\usepackage[scaled=.85]{beramono}
\usepackage[libertine,cmintegrals,cmbraces,vvarbb]{newtxmath} 
\usepackage[scr=boondoxo]{mathalfa}
\usepackage[lf]{carlito}

\newcommand\blfootnote[1]{%
  \begingroup
  \renewcommand\thefootnote{}\footnote{#1}%
  \addtocounter{footnote}{-1}%
  \endgroup
}

\begin{document}

\title{Anionic Character of the Conduction Band of Sodium Chloride}

\author{Christopher C. Leon}
\thanks{These authors contributed equally to this work.}
\blfootnote{This document is the Accepted Manuscript version of a Published Work that appeared in final form in Nature Communications, after peer review and technical editing by the publisher. To access the final edited and published work see \url{https://doi.org/10.1038/s41467-022-28392-8}, {\it Nat. Commun.} {\bf 13}, 981 (2022).\\}
\author{Abhishek Grewal}
\thanks{These authors contributed equally to this work.}
\author{Klaus Kuhnke}
\email{k.kuhnke@fkf.mpg.de}
\affiliation{Max-Planck-Institut f\"ur Festk\"orperforschung, Heisenbergstra{\ss}e 1, 70569 Stuttgart, Germany}
\author{Klaus Kern}
\affiliation{Max-Planck-Institut f\"ur Festk\"orperforschung, Heisenbergstra{\ss}e 1, 70569 Stuttgart, Germany}
\affiliation{Institut de Physique, {\'E}cole Polytechnique F{\'e}d{\'e}rale de Lausanne, 1015 Lausanne, Switzerland}
\author{Olle Gunnarsson}
\email{o.gunnarsson@fkf.mpg.de}
\affiliation{Max-Planck-Institut f\"ur Festk\"orperforschung, Heisenbergstra{\ss}e 1, 70569 Stuttgart, Germany}


\begin{abstract}
The alkali halides are ionic compounds.
Each alkali atom donates an electron to a halogen atom, leading to ions with full shells.
The valence band is mainly located on halogen atoms, while, in a traditional picture, the conduction band is mainly located on alkali atoms.
Scanning tunnelling microscopy of NaCl at 4 K actually shows that the conduction band is located on Cl$^-$ because the strong Madelung potential reverses the order of the Na$^+$ $3s$ and Cl$^-$ $4s$ levels.
We verify this reversal is true for both atomically thin and bulk NaCl, and discuss implications for II-VI and I-VII compounds.
\end{abstract}

\maketitle


\section*{Introduction}
Insulating compounds provide a rich and very varied physics.
One distinguishes between ionic \cite{seitzModernTheory1940}, Slater \cite{Slater}, Mott \cite{Mott}, and charge transfer \cite{Fujimori1, Fujimori2, Sawatzky, Zaanen} insulators, involving quite different mechanisms and having different properties.
Slater and Mott insulators primarily involve a partly filled band, in which antiferromagnetic or Coulomb interactions open up a gap.
The late transition metal oxides are examples of charge transfer insulators, involving the O $2p$-band and the, partly filled, transition metal $3d$ band.
Coulomb interactions are essential also in this case.
Ionic insulators, e.g., alkali halides, appear to be conceptually simpler.
In a one-particle picture, all bands are completely full or empty, and Coulomb interactions are less essential.
There is a large charge transfer from the alkali atoms to the halogen atoms, which become positively and negatively charged ions.
The valence band is primarily of halogen $p$ character.
This band is separated from the conduction band by a large gap.
The conduction band is assumed to have mainly alkali character, involving an empty alkali $s$-level outside a full shell.
While this picture has been presented in many text books and publications \cite{seitzModernTheory1940, harrison1970solid, stonehamTheoryDefects, chingBandTheory1995, elliott1998physics, harrisonElementaryElectronic1999,
grosso2000solid}, this view has been occasionally be questioned on theoretical grounds.
Slater and Shockley \cite{slaterOpticalAbsorption1936} suggested a different picture for NaCl, in which the conduction band also has a substantial Cl $4s$ character.
Clark \cite{clarkCalculationElectronic1968, clarkAugmentedPlane1968}, de Boer and de Groot \cite{deboerOriginConduction1999a, deboerGrainSalt1999}, as well as Olsson \textit{et al.}~\cite{olssonScanningTunneling2005}, performed band structure calculations for NaCl and concluded that the conduction band is actually mainly located on the Cl$^-$ ions.
These theoretical results, however, seem to have been largely overlooked or ignored, possibly because of problems of uniquely assigning charges to ions.

We present a heuristic calculation which suggests that a very large Madelung potential \cite{Madelung1918Das} ($\approx9$ eV) can reverse the order of the Cl$^-$ $4s$ (above the vacuum level) and Na$^+$ $3s$ (at $\approx-5$ eV) levels and imply a conduction band of mainly Cl character.
(The Madelung potential is the potential at any ion position in an ionic crystal due to the combined electrostatic potentials of the infinite number of ions in the crystal.)
However, such a calculation (and similar considerations) alone is not decisive proof of this reality because of confounding factors such as the large spatial extent of s orbitals and non-unique assignments of charges to ions.
For exactly this reason, we perform an experimental study of the conduction band of NaCl using scanning tunnelling microscopy (STM) providing a real space picture of states, which are centred on the Cl$^-$ ions across the entire band gap.

\begin{figure*}[t]
\includegraphics[width=\linewidth]{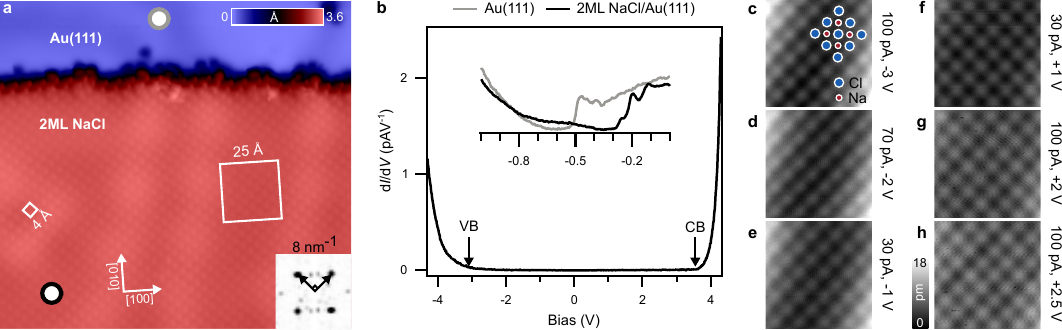}
\caption{{\bf STM of 2 ML NaCl(100) on Au(111) with topography and d\textit{I}/d\textit{V} measurements.} (a) Overview of the measurement area ($150\times140$ \AA$^2$, $I_T=8$ pA, $V_S=-50$ mV). $25\times25$ \AA$^2$ region marked, investigated for bias values between $-3.0$ V and $+2.5$ V. Inset: Fourier transform of (f) centred over a 8 nm$^{-1}\times$8 nm$^{-1}$ region in reciprocal space.
(b) Differential conductance (d$I$/d$V$) measured atop 2 ML NaCl(100)/Au(111) (marked by the black ring in (a)). The arrows point to onset of valence (VB) and conduction band (CB). Inset. Surface state features of Au(111) and related interface state of 2 ML NaCl(100)/Au(111) measured at the positions marked in (a) by a gray ring and a black ring, respectively. Vertical scale: arbitrary units.
(c-h) Raw data showing grayscale topography images obtained for sample bias between $-3.0$ V and $+2.5$ V as indicated. The red and blue dots in (c) mark the positions of Na and Cl ions of the NaCl(100) lattice, respectively. The corresponding tunnelling parameters are indicated next to each topography image.}
\label{fig:1}
\end{figure*}


\section*{Results}
Despite the many STM studies addressing the structure and growth of NaCl films \cite{glocklerInitialGrowth1996, hebenstreitAtomicResolution1999, bennewitzAtomicallyResolved2000, reppRastertunnelmikroskopieUnd2002, lauwaetResolvingAll2012Phys.Rev.B}, such a detailed study of the NaCl conduction band has not yet been performed, possibly because it is very challenging to obtain atomic resolution on NaCl at positive bias (e.g., $U>1$ V) \cite{hebenstreitAtomicResolution1999}.
These difficulties preclude a straightforward spatial mapping of the NaCl conduction band by STM and an identification of the relative local density of states at Na and Cl positions.
The instability of the tunnelling condition necessary to achieve atomic resolution simply prohibits addressing the conduction band directly.

We circumvent this problem by making the bias very positive to approach the conduction band very closely without exceeding the band gap, and harnessing our understanding of the tunnelling process under these conditions.
We present STM images of NaCl(100) on Au(111), varying the bias over a large range.
They show that tunnelling happens through the Cl$^-$ ions, even for energies just below the conduction band, indicating that the conduction band also has mainly Cl character.

To support these conclusion we present tight-binding (TB) calculations for two models of a NaCl film on a Au surface, with the conduction band mainly on either the Na$^+$ or Cl$^-$ ions.
These calculations show that features in STM images for energies just below the conduction band indeed are located on the same ions as the conduction band itself.

We compare NaCl on Au and bulk NaCl theoretically. We find that the order of the Na $3s$ and the Cl $4s$ levels are reversed by similar amounts in both cases, as discussed extensively in Supplementary Note 2.
The conclusions obtained for NaCl on Au should therefore also apply to bulk NaCl.

We furthermore extend the discussion of NaCl to II-VI and other I-VII compounds, suggesting that the conduction band could have a substantial weight on the anion also in these cases.

\subsection*{STM topography of NaCl over the entire band gap}
Because few experimental techniques can characterise the conduction band character of the alkali halides with atomic resolution, we investigated the NaCl(100) surface with STM.
Since directly accessing the conduction band of NaCl(100) involves high electric fields that destroy its structural integrity \cite{hebenstreitAtomicResolution1999}, we do measurements at voltages as close as possible to the conduction band edge while staying within the band gap of NaCl.
Topographs are measured at constant current whose colour scale represent tip-sample distance.
By repeating the measurements at certain voltages, the tip quality was monitored and care was taken that the relatively high bias voltages did not modify the tip character or shift the tip apex. 
Similarly, the tip drift over time, typical for STM, was monitored and corrected in the presented data.

Figure \ref{fig:1}a shows a region of Au(111) covered with a 2 ML thick (apparent height: $309\pm3$~\AA) NaCl terrace, consistent with prior studies \cite{lauwaetDependenceNaCl2012a}.
Atop the NaCl terrace, the protrusions form a square lattice whose unit cell length is 4 \AA~(Fig. \ref{fig:1}a).
The square lattice is reproduced in reciprocal space (Fourier transform inset, Fig. \ref{fig:1}a).

We survey the electronic structure of the system at selected positions marked by gray and black rings in Fig. \ref{fig:1}a.
The pronounced surface state of Au(111) shows an onset at $-500$ mV measured at the gray ring.
This onset shifts to $\approx -250$ mV when measured on Au-NaCl at the black ring. This shift originates from the Pauli interaction experienced between NaCl and Au(111).
NaCl acts as a dielectric, polarises and weakens the image potential arising from Au(111) alone \cite{diekhonerSurfaceStates2003, reppSnellLaw2004} (inset Fig. \ref{fig:1}b) measured at the black ring.
A large range voltage scan of differential conductance indicating density of states of NaCl is shown in Fig. \ref{fig:1}b.
Arrows point to the valence and conduction band onsets at $-3.0$ V and $+3.6$ V.

The region marked by the white box, in the fcc region of herringbone reconstruction of Au(111), in Fig. \ref{fig:1}a is scanned with high stability and signal to noise ratio (Fig. \ref{fig:1}c-h) for applied sample voltages from $-3$ V to $+2.5$ V. respectively.
At applied bias of $-3.0$ V (Fig. \ref{fig:1}c), a voltage close to the valence band edge of NaCl (Fig. \ref{fig:1}b), a square lattice of protrusions is seen whose intensity is proportional to the $z$-corrugation.
Each protrusion is assigned to Cl.
The assignment of protrusions at negative voltage to Cl is in agreement with many earlier studies \cite{hebenstreitAtomicResolution1999, olssonDensityFunctional2003, olssonScanningTunneling2005}.
We do not observe any change in position of $z$-corrugation at negative voltages (Fig. \ref{fig:1}c-e).
These measurements further corroborate the valence band being mainly Cl in character.

\begin{figure*}
\includegraphics[width=\linewidth]{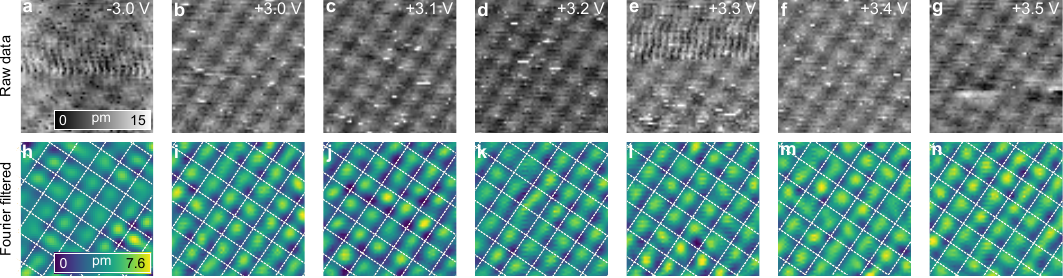}
\caption{{\bf STM of 3 ML NaCl(100) on Au(111) at voltages close to both edges of the band gap.} (a-g) Raw (grayscale) and (h-n) Fourier filtered (coloured) at the indicated voltages, $I_T=15$ pA. Scale: $24\times24$~\AA$^2$.
The white dashed lines mark the position of Na ion rows of the NaCl(100) lattice.}
\label{fig:2}
\end{figure*}

We now consider the same region scanned at positive bias.
Simple but ultimately misleading electrostatic arguments suggest that the topographic protrusions would become centered over Na$^+$ ions.
For instance, electron extraction at negative bias is helped at the Cl$^-$ positions with the Cl$^-$ being relatively electron-rich.
Electron injection at positive bias is helped at the Na$^+$ positions with the Na$^+$ being relatively electron-poor.
Thus, one might anticipate that inverting the applied bias polarity should result in an inversion of the topographic contrast due to $z$-corrugation resulting from Na$^+$ ions instead of Cl$^-$ ions.

We perform this experiment and obtain evidence to the contrary.
Figure \ref{fig:1}f-h show the result of probing the NaCl layer at positive bias up to $+2.5$ V.
By comparing these topographic measurements with those obtained at negative bias (Fig. \ref{fig:1}c-e), it is evident that the protrusions remain centred over Cl and do not shift over to Na.
If the conduction band were cationic in character, one would have expected to see topographic intensity over the Na$^+$ ions.
Yet no contrast inversion is observed when the bias polarity is inverted.
In the band gap of NaCl, the $z$-corrugation is correlated to the electronic states of energetically nearest neighbouring bands.
Thus, the experiment shows that the electronic states of NaCl have substantial weight on Cl rather than Na, irrespective of bias polarity.
When one tunnels at energies within the band gap of NaCl, electron transport in NaCl is dominated by the role of Cl$^-$ ions at both negative and positive bias.
This implies that both the valence and conduction bands of NaCl are anionic in character, a point that is to be emphasised.

The expected role of Na$^+$ ions in the conduction band of NaCl asserted by many textbooks and publications is simply absent.
Rather, we see evidence for Cl functioning as both an electron donor and acceptor in NaCl.
We conceptualise the electronic structure of NaCl as that of spectator Na$^+$ cations holding together Cl$^-$ anions that are mainly responsible for electron conduction.

To further evidence the markedly anionic (rather than cationic) character of the NaCl conduction band, we expand the voltage range of the measurements to $+3.5$ V in exchange for greatly reduced tip stability and signal to noise ratio.
However, we still keep the absolute piezo drift ($0.014$ nm min$^{-1}$) during data acquisition to be significantly smaller than the Cl$^-$--Na$^+$ ion spacing ($\approx 0.3$ nm) and along a direction that avoids introducing spurious contrast inversions due to drift.
These efforts are shown in Fig. \ref{fig:2} in which a different sample area of 3 ML NaCl(100) ($483.5 \pm 5.4$ pm thickness) on Au(111) is scanned at select voltages and analysed with Fourier filtered scans.
As discussed by Lauwaet {\it et al.} \cite{lauwaetDependenceNaCl2012a}, the interface state (see inset Fig. 1b) wave function for NaCl(100) on Au(111) does not extend further than two NaCl layers. In the discussion in Supplementary Note 2, we conclude that electronically, 2 ML and 3 ML NaCl are identical. In contrast to Fig. \ref{fig:1}, the scans in Fig. \ref{fig:2} are taken across the well-defined fcc and hcp regions of the Au(111) herringbone reconstruction

At bias $>+2$ V, the risk of tip changes and spontaneous defect creation in the NaCl layer increase considerably.
As seen in Fig. \ref{fig:2}a-g, while tip transients do prevent smooth scanning at high bias, they do not completely destroy the important atomic resolution.
The low and high frequencies are Fourier filtered to mitigate noise and to clearly highlight the periodic $z$-corrugation.
The measurements in Fig. \ref{fig:2} attain atomic resolution on NaCl even at high bias than the ones in Fig. \ref{fig:1}.
Once again, at $-3$ V, $z$-corrugation is assigned to Cl positions and does not shift to Na positions (lattice positions marked by white dashed line in Fig. \ref{fig:2}h-n) at any of the indicated applied voltages in Fig. \ref{fig:2}.
Repeating these measurements at intermediate voltages between $-3.0$ V to $+3.5$ V reveals no evidence for electron density maxima at Na.
The results in Fig. \ref{fig:2} show that electron density remains centred over Cl, even up to $+3.5$ V (Fig. \ref{fig:2}g, n) which is quite close to the empirically measured conduction band edge ($\approx +3.6$ V) shown in Fig. \ref{fig:1}b.
The second data set thus demonstrates that the conclusions about tunnelling being weighted on Cl in NaCl holds more generally and is observed on different NaCl layer heights, on fcc as well as hcp areas and by using a tip after different preparation.

\subsection*{Tight binding calculations}
We now perform tight-binding calculations for a model of a three-layer NaCl(100) film on a Au(111) surface (see Supplementary Note 1).
We use two different sets of parameters for the NaCl film, resulting in a conduction band mainly of either Na $3s$ or Cl $4s$ character.
In the first set we use $3s$ and $3p$ basis states on the Cl atoms and in the second set the Cl $3s$ basis states are replaced by $4s$ basis states.
States in the energy gap of NaCl are described as linear combinations of valence and conduction states.

\begin{figure*}[t]
\includegraphics{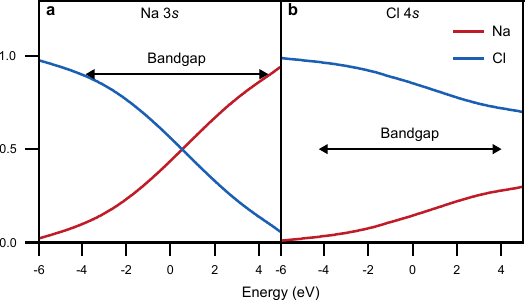}
\caption{{\bf Character of the gap states of NaCl including Madelung potential.}
Normalised relative weight of gap states on Na and Cl atoms in the outermost NaCl layer as a function of energy, $E$, in or close to the gap (indicated by the arrows).
The parameters were adjusted so that the conduction states have mainly (a) Na character (b) Cl character.
The Figure illustrates that electrons travelling through the gap at energies close to the conduction band, have similar character as those in the conduction band.}
\label{fig:3}
\end{figure*}

In Fig. \ref{fig:3}a, we use conventional parameters, including $3s$ and $3p$ states on the Cl atoms \cite{harrisonElementaryElectronic1999} and neglect the Madelung potential. 
(Note, that in the calculation with Cl $3s$ and $3p$ orbitals only, the conduction band is repelled upwards by these states.
Including higher states on Cl would tend to have the opposite effect, requiring a higher Na $3s$ level to obtain the correct band gap. 
This may then require a contribution from the Madelung potential also in this case.)
As we increase the energy through the gap, the character of the state changes from mainly Cl character close to the valence band to mainly Na character close to the conduction band, in contradiction to the experimental finding.

Figure \ref{fig:3}b shows results for the second set of parameters, including $3p$ and $4s$ states on the Cl atoms.
In this case, it is essential to include a Madelung potential to obtain a conduction band of mainly Cl character and a correct band gap.
The result is that the gap states have mainly Cl character throughout the whole gap, in agreement with experiment.
This then provides further evidence that not only the gap states but also the conduction band has mainly Cl character.
We emphasize again, the inclusion of Cl $4s$ states in these calculations.

Tunnelling through an insulator band gap proceeds via the electronic states of the energetically nearest neighbouring bands.
If NaCl were composed of a Cl$^-$ based valence band and a Na$^+$ based conduction band, an atomically resolved STM map of it would show voltage-dependent contrast that inverts at a specific voltage in the band gap suggested in Fig. \ref{fig:3}a.
However, this inversion does not happen in the experiment.
In fact, the same type of ion appears bright at energies near both the valence and conduction bands, corroborating the picture that both bands have most of their weight on the same type of ion, specifically Cl$^-$ in the case of NaCl (Fig. \ref{fig:3}b).

\section*{Discussion}
Naively applying classical electrostatics, one would predict that the negatively charged electrons avoid the negatively charged Cl$^-$ in favour of the positively charged Na$^+$.
Moreover, because Cl$^-$ has a full shell electron configuration [Ne]3s$^2$3p$^6$, the extra tunnelling electron must be accommodated in the higher energy Cl $4s$ orbital, which also appears to impose an insurmountable energetic cost.

To understand why the Cl $4s$ states might, nevertheless, be so important, we make a few very simple considerations.
For fully ionised atoms (in the sense of Cl$^-$ and Na$^+$ having exact full shell electronic configurations), the Madelung potential is 8.9 eV, leading to a strong upward shift of the Na $3s$ level, which for a neutral atom is just at $-5.1$ eV \cite{mooreIonizationPotentials1970}.
Even if the $3s$ orbital relaxes somewhat, it may then be pushed above the vacuum level by the Madelung potential.

Meanwhile, the $4s$ level of a free Cl$^-$ ion is above the vacuum level.
The issue is whether the Madelung potential can pull this level below the Na $3s$ level.
To puzzle this out, we use a thought experiment to increase the nuclear charge of Cl by two.
This turns Cl$^-$ into K$^+$, for which the $4s$ level is at $-4.1$ eV \cite{mooreIonizationPotentials1970}, below the (raised) Na $3s$ level.
The question is if the Madelung potential has a similar attractive effect on the Cl $4s$ level.
We use Slater's rules \cite{slaterAtomicShielding1930} to approximate the $4s$ state of K$^+$ as $\phi_{4s}\sim r^{2.7}\exp{(-2.2r/3.7)}$, where $2.2$ is the effective nuclear charge and $n=3.7$ is the effective principal quantum number.
We then calculate the attractive potential on the $4s$ state due to increasing the nuclear charge of Cl by $\Delta Z=2$.
The magnitude of this potential is indeed comparable to the Madelung potential of 8.9 eV.
In reality, the atoms are not fully ionised and so the Madelung potential is smaller.
Simple estimates suggest, however, that this effect and several other smaller effects do not change the picture in an essential way (see Supplementary Note 3).
The K $4s$ orbital is very extended and complicates the interpretation of these results.
Nevertheless, these simple considerations make it plausible that the Madelung potential can reverse the order of the Cl$^-$ $4s$ level and the Na$^+$ $3s$ level, resulting in a conduction band of mainly Cl character in NaCl.
This is confirmed by the STM measurements in Figs. \ref{fig:1} and \ref{fig:2} that show well-defined features centred on the Cl ions.
This interpretation is robust because the STM measurements are performed in real space and do not depend on interpretations expressed in terms of orbitals.

We now discuss differences between bulk NaCl and NaCl on Au, using tight-binding calculations described in the Supplementary Note 2.
For a NaCl film on Au, we find an appreciable charge transfer to the Au substrate, consistent with an observed substantial reduction of the work function.
This raises the potential of the outer layers of the film substantially, but the effect on the potential difference between the Na and Cl sites in the outermost layer is very small.
The assumed smaller lattice parameter of the NaCl film on Au increases the Madelung potential, while the finite thickness of the film reduces it.
Adding up all effects, we find that the tendency to put the conduction band on the Cl atoms should be comparable (difference on the order of $0.1$ eV, see Supplementary Table IV) for bulk NaCl and for NaCl on Au, implying that the effects observed here should apply also to bulk NaCl.

\section*{Conclusion}
In nanoscience, a NaCl film is often used as a buffer between a substrate and a molecule to be studied \cite{reppMoleculesInsulating2005, liljerothCurrentInducedHydrogen2007a}.
Due to the large NaCl gap, the applied voltage is typically such that the electrons tunnel through the NaCl gap.
The way the substrate couples to the molecule then depends on whether the corresponding electrons mainly tunnel through the Cl$^-$ or Na$^+$ ions.
The results above should then be important for interpreting such experiments.

In contrast to the discussion above, in nanoscience it is often assumed that the  electrons tunnel through a band gap via some state, not well specified, being different from both the valence and conduction states \cite{chenSpinTripletMediatedUpConversion2019a}.
To address this issue, we could have included additional basis states, which would then have to be orthogonal to the valence and conduction states of the NaCl film.
But such states would necessarily be higher in energy than the conduction states already included.
This implies that the additional basis states would not be very important for describing gap states, which are located below the conduction band.
By contrast, the explicit use of just valence and conduction states as basis states should already give a good qualitative picture of what happens when tunnelling through the gap, and such calculations are relatively easy to interpret.

In view of these results for Na$^{+}$Cl$^{-}$ it is interesting to discuss Mg$^{2+}$O$^{2-}$.
In analogy to the discussion above, we do the thought experiment of increasing the nuclear charge of O$^{2-}$ by three, obtaining Na$^{+}$ with its 3s level at $-5.1$ eV.
The attractive interaction of this additional nuclear charge with the $3s$ level is $20.0$ eV, comparable to the Madelung potential $23.9$ eV for the ions Mg$^{2+}$ and O$^{2-}$.
This suggests that the conduction band of MgO could be located on oxygen, involving mainly O $3s$ levels.
Indeed, this is what was found in band structure calculations by de Boer and de Groot \cite{deboerConductionBand1999}.
It would then be very interesting to experimentally check the character of the conduction bands both for the II-VI and for other I-VII ionic compounds.

Beyond these simple salts, it would indeed be fruitful to further experiment with materials whose valence and conduction bands take on unusual combinations of cationic and anionic character, e.g., that of the valence band being cationic \cite{blasseMaterialsCationic1981}, or that of the conduction band being anionic \cite{zhuravlevNatureElectronic2009}.
There may also be complementary mechanisms to probe that induce electronic structure rearrangements like that of the Madelung potential.

To summarise, using STM we have shown that the conduction band of NaCl has mainly Cl character, contrary to widespread belief.
That the conduction band can be anionic in character is related to the large Madelung potential, which pushes the Na $3s$ level upwards and pulls the Cl $4s$ level downwards, which for a free Cl$^-$ species is above the vacuum level.
We provided a back of the envelope calculation to make it plausible that the Madelung potential could actually reverse the order of the Cl$^-$ $4s$ and the Na$^+$ $3s$ levels, leading to this result. This electronic structure may couple into other physical processes.
Incidentally, it is the negative ion vacancies that are important to electrical conduction near the melting point of some alkali halides \cite{raoElectricalConductivity1975}.

\section*{Methods}
\setlength\parindent{0pt}\textbf{Sample preparation and details of STM measurements}  -- The experiments were carried out with a home-built low-temperature STM operated at $T=4.3$ K, UHV ($<10^{-11}$ mbar) \cite{kuhnkeVersatileOptical2010}.
The Au(111) single-crystal ($>99.999\%$) sample was cleaned by repeated cycles of Ar$^+$ ion sputtering at $10^{-6} mbar$ range argon pressure with 600 eV acceleration energy and subsequent annealing to 873 K.
The sample heating and cooling rate was about 1 K/s.
NaCl was evaporated thermally from a Knudsen cell held at $900$ K, with the Au(111) surface held at $300$ K, to obtain defect-free, (100)-terminated NaCl islands.
An electrochemically etched gold wire \cite{yangFabricationSharp2018} ($99.95\%$ purity) was used as tip in the experiment.
To ensure a metallic tip, the Au wire is further prepared by controlled tip indentations ($1-3$ nm, $V=50-100$ mV) in Au(111) until atomic resolution is obtained at tunnelling current set point: $I_T=10$ pA, $+1$ V.
The text always specifies bias voltages of the metal substrate with respect to the grounded tip.
Differential conductance (d$I$/d$V$) spectra were measured using a standard lock-in technique with a bias modulation of $V_{rms}=4$ mV (Fig. \ref{fig:1}b inset) and $V_{rms}=10$ mV (Fig. \ref{fig:1}b) at 629 Hz.
Scanning tunnelling microscopy/spectroscopy data were analysed using self-written MATLAB code.

\end{document}


\title{Supplementary Information for\\ Anionic Character of the Conduction Band of Sodium Chloride}
\maketitle

\section*{Supplementary Note 1: Parameters}
Below we describe the parameters used to calculate the gap states in the NaCl film for the two cases when the conduction band is mainly on the Na atoms or mainly on the Cl atoms.
We consider three NaCl(100) layers on five Au(111) layers, using periodic boundary conditions parallel to the surface.

\subsection*{Au}
To describe Au, the hopping parameters of Harrison \cite{harrisonElementaryElectronic1999} were used.
We also used the Harrison level energies $\varepsilon_{6s}=-6.98$ eV  and $\varepsilon_{5d}=-17.78$ eV as a starting point.
A $6p$ level at $\varepsilon_{6p}=-2.8$ eV was added.
We then required that the top (${\bar M}$-point) of the $5d$ band is placed at 1.7 eV below the Fermi energy \cite{sheverdyaevaEnergymomentumMapping2016}. This required a shift of the $5d$ level to $\varepsilon_{5d}=-14.33$ eV.
These parameters are summarised in Supplementary Table~\ref{table:1}.

\subsection*{NaCl}
To describe NaCl, we also used the hopping parameters of Harrison \cite{harrisonElementaryElectronic1999}.
Two cases were considered.
First we followed Harrison \cite{harrisonElementaryElectronic1999} and used the parameters $\varepsilon_{{\rm Na}, 3s}=-4.96$ eV, $\varepsilon_{{\rm Cl}, 3s}=-29.2$ eV and $\varepsilon_{{\rm Cl},3p}=-13.78$ eV.
A Na $3p$ state was also added ($\varepsilon_{{\rm Na},3p}=-1$ eV).
The effects of the Madelung potential were neglected.
This led to a somewhat too large gap compared with the experimental result 8.5 eV \cite{pooleElectronicBand1975}.
The Na $3s$ level was therefore lowered somewhat to $\varepsilon_{{\rm Na}, 3s}=-5.55$ eV.
These parameters are shown in Supplementary Table~\ref{table:1} as case 1.

Alternatively, we followed de~Boer and de~Groot \cite{deboerOriginConduction1999a} and assumed that the conduction band has Cl
$4s$ character.
For this purpose we replaced the Cl $3s$ level by a $4s$ level at $\varepsilon_{{\rm Cl},4s}=1.35$ eV, leaving the Cl $3p$ level at $\varepsilon_{{\rm Cl},3p}=-13.78$ eV.
We then took the Madelung potential into account and shifted the Na levels upwards ($\varepsilon_{{\rm Na},4s}=4$ eV and $\varepsilon_{{\rm Na},3p}=8$ eV).
This also resulted in a band gap of 8.5 eV \cite{pooleElectronicBand1975}.
These parameters are shown in Supplementary Table~\ref{table:1} as case 2.

\subsection*{Combined system}

For the hopping between the Au and NaCl slabs we introduced a cut off at a distance of
\begin{equation*}
\sqrt{d_{\rm Au-NaCl}^2+a_{\rm NaCl}^2/2}.
\end{equation*}
The hopping integrals were then calculated according to the prescription of Harrison \cite{harrisonElementaryElectronic1999}.
Then each Au atom in the outermost Au layer typically couples to about 5-7 atoms in the closest NaCl layer.
We then shifted all the Au level energies so that the Fermi energy is at zero. Calculations using the GW method \cite{hedinNewMethod1965} find that for NaCl on Au, the top of the NaCl valence band is about 5 eV below the Fermi energy of Au \cite{wangQuantumDots2017}.
We then shifted all NaCl levels correspondingly.

Calculations find that the lattice parameter of a three layer NaCl film on Au is reduced compared with bulk NaCl \cite{chenPropertiesTwodimensional2014}.
We therefore used the calculated reduced lattice parameter $a_{\rm NaCl}=5.54$ \AA \ \cite{chenPropertiesTwodimensional2014}.
We used the calculated separation $d_{\rm Au-NaCl}=3.12$ \AA \ between the Au surface and the NaCl film \cite{chenPropertiesTwodimensional2014}.
For Au we use the lattice parameter $a_{\rm Au}=4.07$ \AA \ \cite{daveyPrecisionMeasurements1925}.

\begin{table}
\captionsetup{name=SUPPLEMENTARY TABLE}
\caption{Level positions used for NaCl on Au.
Case 1 and 2 correspond to a conduction band of mainly Na $3s$ or Cl $4s$ character, respectively.
}\label{table:1}
\begin{tabular}{lcrrr}
\hline
\hline
Element & Case & $s$  & $p$ & $d$ \\
\hline
Au (6s, 6p, 5d)  & 1, 2   &  -6.98  &  -2.8 & -14.33 \\
Na (3s, 3p) &  1 & -5.55 & -1.0 &  - \\
Cl (3s, 3p) &  1 & -29.2 & -13.78 & - \\
Na (3s, 3p) &  2 &  4    &   8  & -  \\
Cl (4s, 3p) &  2 &  1.35 & -13.78 & - \\
\hline
\hline
\end{tabular}
\end{table}

\section*{Supplementary Note 2: Comparison of a thin NaCl film on Au and bulk NaCl}

{\it Bulk NaCl}: In the main text, we used the Madelung potential for bulk NaCl and fully ionised atoms. This potential is given by
\begin{equation}
V = 1.7476 \; { \frac{e^2}{ 4 \pi \epsilon_{0} d} }=8.92~{\rm V},
\end{equation}
where $d=a_{\rm NaCl}/2=2.82$ \AA \ is the separation of the nearest neighbour Cl and Na atoms, $\epsilon_{0}$ is the vacuum electric permeability and we assume full positive and negative elementary charges on Na and Cl, respectively.
We now discuss the corrections to this, focusing on the differences between bulk NaCl and a thin NaCl film on Au.

Tight-binding calculations for bulk NaCl give the net charge 0.805 of the ions (see Supplementary Table~\ref{table:0}).
This reduces the Madelung potential to 7.18 V.
This is the reference for the following considerations.
We now focus on the differences between bulk NaCl and a few layers of NaCl on an Au substrate.

\begin{table}[b]
\captionsetup{name=SUPPLEMENTARY TABLE}
	\caption{Orbital weights and net charges on Na and Cl atoms for bulk NaCl.
		\label{table:0}}
	\begin{tabular}{cccccc}
		\hline
		\hline
		 Na $3s$  & Na $3p$ & Net charge Na   & Cl $4s$ & Cl $3p$ & Net charge Cl \\
		\hline
		0.069 & 0.126 & 0.805& 0.004  & 5.801& -0.805 \\
		\hline
		\hline
	\end{tabular}
\end{table}

{\it Reduced lattice parameter}:
The lattice parameter of a NaCl film is assumed to be reduced  to $a_{\rm NaCl}=5.54$ \AA \ \cite{chenPropertiesTwodimensional2014}.
This increases the Madelung potential by 0.129 V compared with the bulk.
This is shown in Supplementary Table~\ref{table:n2} as a positive contribution under ``Film lattice parameter smaller''.

{\it Film thickness}:
Due to the NaCl film being just a few layers, the Madelung potential of the film is reduced by about 0.27 V compared with the bulk.
This is shown under ``Finite film thickness'' in Supplementary Table~{\ref{table:n2}.

\begin{table}
\captionsetup{name=SUPPLEMENTARY TABLE}

	\caption{Net total charges on Na and Cl atoms in different layers of a film with 2-4 layers outside an Au substrate and net charge on the Au substrate per pair of Na and Cl atoms.
		\label{table:n1}}
	\begin{tabular}{cccccccccc}
		\hline
		\hline
	        Layer    & \multicolumn{2}{c} {1} & \multicolumn{2}{c} {2}
		    & \multicolumn{2}{c} {3}   & \multicolumn{2}{c} {4}  & Au \\
		\hline
		Layers    & Na  & Cl   & Na  & Cl & Na  & Cl  & Na  & Cl  & Subst. \\
		\hline
		2     & .767 & -.660 & .834 &-.834 &&  &&&-.107 \\
		3     &  .769&-.660&.800&-.805&.836 &-.834  &
		&& -.106 \\
		4      &  .769&-.660&.802&-.805&.802 & -.805 & .836 &-.834 &-.105  \\
		Bulk  & .805 &-.805 &&&&  &&\\
		\hline
		\hline
	\end{tabular}
\end{table}

{\it Charge densities and work function:}
We perform tight-binding calculations for a thin film of a few layers of NaCl on an Au substrate with three layers. Each layer has 324 atoms.
The charges on the atoms in the layers are shown in Supplementary Table~\ref{table:n1}.
For 2-4 NaCl layers there is a net positive charge on the film of about 0.1 electrons per pair of Na and Cl atoms.
Putting the corresponding charge at the centre of the outermost Au layer leads to a large reduction of the work function by about 2.0 eV.
Experimentally, the work function of Au is 5.33 eV \cite{work1}, 5.3-5.6 eV \cite{work2} or 5.5 eV \cite{work3}.
For NaCl on Au(111) the work function is reduced to 4.3 eV \cite{ZheLi}.
Using $I(z)$ spectroscopy, we measure a work function reduction of 1.05 eV  (see Supplementary Figure \ref{fig:s1}) also suggesting a work function of $\approx 4.3$ eV for NaCl on Au(111).
This reduction is not quite as large as in the calculation. The deviation is not surprising, given that the tight-binding calculation only provides charges on atoms but no distortions of wave functions.

{\it Increased charge on outermost layer:}
The hopping between a Cl atom and its neighbouring Na atoms reduces the charge on Cl.
This results in the occupancy of the $3p$ level of 5.801 for bulk NaCl, i.e., a reduction by 0.199 from the nominal value 6.0.
In the outermost layer of a NaCl film, the number of Na neighbours is reduced from six to five.
We then expect fewer Cl $3p$ holes than in the bulk, i.e., the number of holes to be reduced from 0.199 in the bulk to $0.199\times(5/6)=0.166$ in the film, about 0.033 smaller than in the bulk.
This agrees fairly well with the calculated reduction of 0.029 for a two layer film.
The Madelung potential for one single layer of fully ionised atoms is 8.39 V.
The increase of the Madelung potential in the outermost layer due to the extra charge is then $0.029\times8.39=0.243$ V.
For a three layer film we obtain the result $((0.836+0.834)/2-0.805)\times8.39=0.252$ V.
This is shown under ``Increase of charge on the outermost layer'' in Supplementary Table~\ref{table:n2}.

{\it Reduced charge on inner layers:}
We first neglect the charge transfer to the Au substrate and temporarily put this charge on the Na atoms in the layer closest to Au.
The charge transfer between Na and Cl is still slightly smaller on the inner layer than in bulk NaCl.
This slightly reduces the Madelung potential on the atoms in the outermost layer, shown under ``Change charge inner layers''.

{\it Charge transfer to Au:}
The charge transfer to Au from NaCl has a very large effect on the work function (1 eV or more).
However, the work function refers to a uniform shift of the potential well outside the NaCl layers.
Here we are interested in the difference in potential between Na and Cl atoms in the surface layer.
This turns out to be very small for two layers and negligible for three layers (see ``Charge transfer to Au''), although there is a large shift of the potential in the outermost layers.

\begin{figure}[t]
\captionsetup{name=SUPPLEMENTARY FIGURE}
  \caption{\label{fig:s1} Current vs distance measurements with the same tip on Au(111) and on a nearby 2 ML NaCl/Au(111). The measurements are performed at a bias voltage of -0.2 V. Forward and backward scans were averaged using the geometric average. The decay constants for $k_{\text{Au}}$ and $k_{\text{NaCl}}$ convert to a work function difference ($\Delta\phi$) as: $\Delta\phi = 0.01905 (1/k_{\text{NaCl}}^2 - 1/k_{\text{Au}}^2) ($nm$^2$eV).}
  \includegraphics[width=\linewidth]{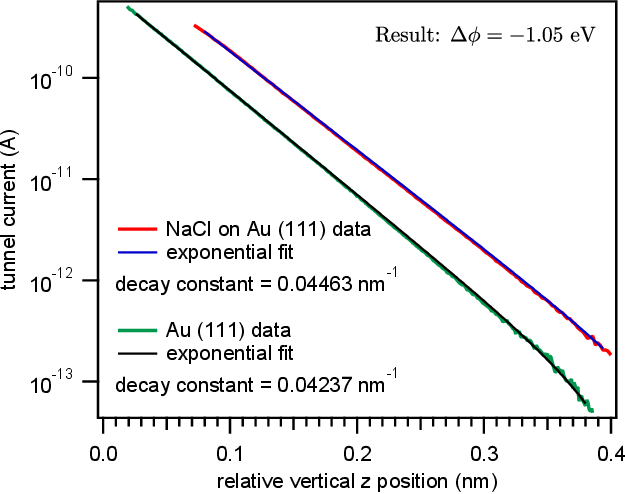}
\end{figure}

{\it Cl $3p-4s$ interaction:}
The increase in the $3p$ charge of the Cl atom at the surface of the film tends to raise the $4s$ level via the on-site Coulomb interaction.
To address this, we use the similarity of a K$^+$ ion and a Cl$^{-}$ ion in the Madelung potential, discussed in the main text.
We use K$^+$ atomic data \cite{Moore} to estimate the shift of the $4s$ level due to the extra $3p$ charge (0.029).
This gives an upward shift of $0.029\times 7.32=0.212$ eV, where 7.32 eV is the $3p-4s$ direct Coulomb integral, derived from atomic data.
We assume a similar but opposite effect on the Na $3s$ level.
This reduces the Na $3s$ and Cl $4s$ splitting, shown as ``On-site Cl contribution'' in Supplementary Table~\ref{table:n2}.
In the end, the increase of the charges in the outermost layer plays a rather small role, due to the approximate cancelling of the on-site Coulomb interaction and the contribution to the Madelung potential.

We conclude that the lowering of the Cl $4s$ level relative to the Na $3s$ level is very similar in bulk NaCl and NaCl on an Au surface, and the effect may even be larger in bulk NaCl.
The observation that the conduction band is mainly on the Cl atoms should then also apply to bulk NaCl.

\begin{table}
\captionsetup{name=SUPPLEMENTARY TABLE}
\caption{Contributions to the difference of the Madelung potential (in $V$) for the outermost layer of a thin NaCl film outside an Au substrate compared with bulk NaCl.
Positive values imply a larger Madelung potential for the NaCl film.
We have also added the contribution from the on-site Coulomb integral due to the increased $3p$ charge on Cl atoms in the outermost layer.
The negative total contribution implies that the Madelung potential is slightly larger in bulk NaCl.\label{table:n2}}
	\begin{tabular}{lrr}
		\hline
		\hline
		Effect     & 2 layers & 3 layers \\
		\hline
		Film lattice parameter smaller            & 0.129 & 0.129\\
		Finite film thickness                     &-0.276 & -0.279\\
		Increase of charge on the outermost layer & 0.244& 0.252\\
		Change charge inner layers                & -0.051&-0.001\\
		Charge transfer to Au                     & 0.019& 0.000\\
		On-site Cl contribution                 &  -0.212&  -0.212\\
		\hline
		Sum                                      & -0.147& -0.111\\
		\hline
		\hline
	\end{tabular}
\end{table}

\section*{Supplementary Note 3: I-VII and II-VI compounds}

Here we discuss the possibilities that I-VII and II-VI compounds in general may have the conduction band on the anion.
We calculate the Madelung potential for fully ionized atoms.
The effect of the Madelung potential on an $s$-orbital outside a free, negatively charged, anion is not trivial to estimate, since for the free ion this orbital is not bound.
Therefore we make the thought experiment of increasing the nuclear charge on the anion by $\Delta Z=2$ for the I-VII compounds and by $\Delta Z=3$ for the II-VI compounds, converting, e.g., Cl$^-$ into K$^+$ or O$^{--}$ into Na$^+$. This results in a bound $s$ orbital outside a full shell.
We can then easily estimate the expectation value of the potential from the extra nuclear charge for the bound $s$-orbital, e.g, the $4s$ orbital of K$^+$.
This provides an estimate of the strength of the potential needed to bind the anion $s$-orbital.
If the Madelung potential is comparable or larger, it seems plausible that it could have pulled the previously unbound $s$-orbital below the vacuum level.
We calculate
\begin{equation}\label{eq:3}
\left\langle\phi(r)\left|\frac{1}{r}\right|\phi(r)\right\rangle,
\end{equation}
where
\begin{equation}\label{eq:4}
	\phi(r)\sim r^{n_{\rm eff}-1}e^{-(Z_{\rm eff}/n_{\rm eff})r},
\end{equation}
is the Slater orbital \cite{slaterAtomicShielding1930} of the alkali ion (e.g., K$^+$ for the case of NaCl) obtained by increasing the nuclear charge by $\Delta Z$, and $Z_{\rm eff}=2.2$ is the Slater effective nuclear charge for that orbital.
The results are shown in Supplementary Table~\ref{table:2}.
We then compare this integral times $\Delta Z$ with the Madelung potential for fully ionised atoms in Supplementary Table~\ref{table:3} for a few ionic compounds.
The Madelung potential is typically comparable to $\langle\phi(r)|{\Delta Z/ r}|\phi(r)\rangle$, which suggests that the conduction band could be on the anion.

\begin{table}[b]
\captionsetup{name=SUPPLEMENTARY TABLE}
\caption{Anions and and the alkali ion obtained by increasing the nuclear charge by $\Delta Z$, the effective quantum number $n_{\rm eff}$ of the alkali ion as well as $\langle \phi|1/r| \phi \rangle$, where $\phi$ is the alkali ion $s$-function.
\label{table:2}}
\begin{tabular}{lcccc}
\hline
\hline
Anions & Alkali ion & $n_{\rm eff}$ & $\langle \phi|1/r| \phi \rangle$ [eV]\\
\hline
O$^{--}$, F$^-$   & Na$^+$ &   3    &  6.7    \\
S$^{--}$, Cl$^-$  & K$^+$  &   3.7  &  4.4    \\
Se$^{--}$, Br$^-$ & Rb$^+$ &  4.0   &  3.7  \\
Te$^{--}$, I$^-$  & Cs$^+$ &  4.2   &  3.4 \\
\hline
\hline
\end{tabular}
\end{table}

\begin{table}
\captionsetup{name=SUPPLEMENTARY TABLE}
\caption{Some ionic compounds and the $\Delta Z$ needed to convert the anion into an alkali ion.
The table also shows the Madelung potential for fully ionised atoms and the attraction on the outermost anion $s$-orbital from the increase, $\Delta Z$, in the nuclear charge.
If these two quantities are comparable, it becomes plausible that the conduction band could be on the anion.
\label{table:3}}
	\begin{tabular}{cccc}
\hline
\hline
Compound & $\Delta Z$ & Madelung [eV] &  $\langle \phi|\Delta Z/r| \phi \rangle$ [eV] \\
\hline
NaCl     &    2 &    8.9             &  8.8  \\
LiI      &    2 &   8.3    &  6.8  \\
RbF      &    2 &   8.9    &  13.4  \\
MgO      &    3 &  23.9    &  20.1  \\
BaO      &    3 &  18.2  &  20.1  \\
\hline
\hline
\end{tabular}
\end{table}

We should, however, remember that the ions are not fully ionised, reducing the Madelung potential.
The $s$-orbital outside the negatively charge anion is also quite extended, making the assignment of its charge to the anion somewhat questionable.
On the other hand, the strong variation of the Madelung potential between the ions may be sufficient to also localise the anion $s$-orbital.
The $s$-orbital on the cation is moved upwards, and it is sufficient that the $s$-orbital on the cation is well above the $s$-orbital outside the anion to obtain these results.
The considerations above, however, can only be seen as suggestive and used to rationalise the rather surprising experimental results for NaCl and, possibly, for other ionic compounds.
Decisive is that STM provides a real space measurement, which does not depend on a discussion in terms of orbitals and that STM for NaCl provides a well-defined image localised on the Cl atoms.\\

\begin{figure}[!h]
\captionsetup{name=SUPPLEMENTARY FIGURE}
  \caption{\label{fig:s2} Evaluation of drift for data in Fig. 2 in the main manuscript. $2.4 \times 2.4$ nm$^2$ sized topographs are recorded at resolution of $64\times64$ pixels. Total data acquisition time for the data set is 10 mins. From linear fit we get drift along x-axis: -2 pixels and along y-axis: 3 pixels. Total drift: 135 pm.}
  \includegraphics[width=\linewidth]{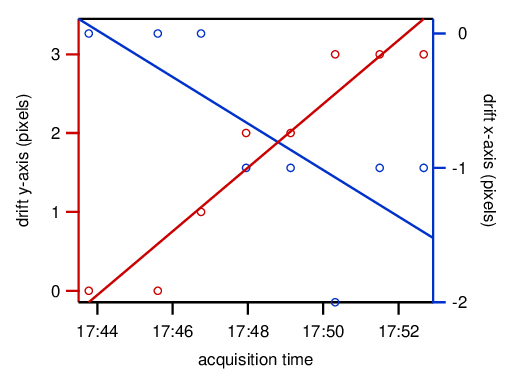}
\end{figure}